\newcommand{\be}{\begin{equation}}
\newcommand{\ee}{\end{equation}}
\newcommand{\ber}{\begin{eqnarray}}
\newcommand{\eer}{\end{eqnarray}}

\newcommand\+{\dagger}

\newcommand\ket[1]{|{#1}\rangle}

\newcommand{\<}{\langle} 
\renewcommand{\>}{\rangle}

\newcommand{\xx}{\textrm{\textbf{x}}}
\newcommand{\yy}{\textrm{\textbf{y}}}

\newcommand{\dsp}{\displaystyle}

\documentclass[twocolumn,showpacs,preprintnumbers,amsmath,amssymb]{revtex4} 

\usepackage{psfrag}
\usepackage{epsfig}
\usepackage{color}

\begin{document}
\tighten
\bigskip
\title{Electronic structure of semiconductor nanoparticles from stochastic evaluation of imaginary-time path integral}
\author {Andrei Kryjevski\footnote{andrei.kryjevski@ndsu.edu}
 } 
\affiliation{Department of Physics,~North Dakota State University,~Fargo,~ND~58108,~USA}
\author{Thomas Luu\footnote{t.luu@fz.juelich.de}}
\affiliation{Institut f\"{u}r Kernphysik \& Institute for Advanced Simulation,
    Forschungszentrum J\"{u}lich, 54245 J\"{u}lich Germany}
\author{Valentin Karasiev\footnote{vkarasev@lle.rochester.edu}}
\affiliation{Laboratory for Laser Energetics,~University of Rochester,~NY~14623,~USA}

\date{\today}

\begin{abstract}
In the Kohn-Sham orbital basis imaginary-time path integral for electrons in 
a semiconductor nanoparticle has a mild Fermion sign problem and is 
amenable to evaluation by the standard stochastic methods. This is evidenced by 
the simulations of silicon hydrogen-passivated nanocrystals, such as $Si_{35}H_{36},~Si_{87}H_{76},~Si_{147}H_{100}$ and $Si_{293}H_{172},$ which contain $176$ to $1344$ valence electrons and range in size $1.0 - 2.4~nm$, utilizing the output of 
density functional theory simulations.
We find that approximating Fermion action with just the leading order 
polarization term results in a positive-definite integrand in the 
functional integral, and that it is a good approximation of the full 
action. We compute imaginary-time electron propagators in these 
nanocrystals and extract the energies of low-lying electron and hole levels. 
Our quasiparticle gap predictions agree with the results of high-precision 
calculations using $G_0W_0$ technique. This formalism can be extended 
to calculations of more complex excited states, such as excitons and trions.
\end{abstract}


\maketitle

Applications of semiconductor nanomaterials require quantitative 
understanding of their electronic structure, including excited state 
properties. In recent years computational studies of atomistic models
of these systems using {\it ab initio} electronic structure techniques 
have proven to be an attractive alternative to actual experiments 
as the ability to explore the vast set of possible configurations is 
inevitably limited.
Currently, Density Functional Theory (DFT) \cite{PhysRev.136.B864,PhysRev.140.A1133} is 
the most useful first-principles atomistic tool for electronic 
structure. It combines reasonable accuracy and applicability, and 
naturally allows inclusion of surfaces, interfaces, dopants, ligands, {\it etc.} 

However, 
DFT 
predicts ground state properties. Therefore, alternative methods are required 
to study the excited states. Currently, 
the most efficient comprehensive {\it ab~initio} approach is  
based on many-body perturbation theory (MBPT), where DFT is augmented by 
the methods of perturbative many-body quantum mechanics. For instance, the GW method 
is used to compute single-particle energies,
the Bethe-Salpeter equation (BSE) is solved for exciton states; 
an MBPT technique for three-body states, such 
as trions, has also been developed~
\cite{PhysRev.139.A796,PhysRevB.34.5390,PhysRevB.62.4927,RevModPhys.74.601,PhysRevLett.116.196804}. 
The resulting energies and wave functions
are subsequently used in calculations of various 
excited-state properties ({\it e.g}, \cite{PhysRevLett.80.3320,PhysRevLett.92.077402,doi:10.1021/acs.jpclett.8b02288}). 

Non-perturbative high-precision quantum Monte Carlo (QMC) techniques, 
such as fixed node, diffusion, and auxiliary field, 
exist (see, {\it e.g.},
\cite{PhysRevLett.90.136401,doi:10.1021/cr2001564,Zhang2018}).
Importantly, in these electronic systems the Fermion sign problem \cite{SK,PhysRevB.41.9301} 
is mild enough to allow precise simulations. 
However, these MC methods are mostly suited for studying the ground state. 
Work to develop a QMC technique for excited states has started, but the 
studies so far have been based on the tight-binding approximation and applied 
to model systems (see, {\it e.g.}, \cite{PhysRevLett.83.2777,PhysRevLett.82.4155}) and to graphene and carbon nanotubes 
\cite{PhysRevB.79.165425,Smith:2014tha,Ulybyshev:2013swa,Otsuka:2015iba,Beyl:2017kwp,PhysRevB.84.075123,PhysRevB.93.155106}. 

Here we present a DFT-based comprehensive nonperturbative QMC 
technique for 
a semiconductor nanoparticle, where 
excited states, such as electrons and holes, 
excitons and trions, can be obtained from the 
output of the same MC simulation. The system-specific Kohn-Sham (KS) 
orbitals are used as a basis in the electron action in the path integral 
representation of the statistical sum. Our results suggest that in this approach there 
is only a mild Fermion sign problem and evaluation by the standard stochastic 
importance sampling methods employed in, {\it e.g.}, lattice Quantum 
Chromodynamics (QCD) (see \cite{gattringer2009quantum} and references within) is possible. 
Specifically, we present results of simulations of several semiconductor 
nanocrystals, such as $Si_{35}H_{36},~Si_{87}H_{76},~Si_{147}H_{100}$ and $Si_{293}H_{172},$ including low-lying single-particle energies. 

In the Born-Oppenheimer approximation the non-relativistic Hamiltonian for valence electrons is
\ber
{\rm H}&=&\int{\rm d}{\bf x}~\psi^{\+}_{\alpha}\left(-\frac{\hbar^{2}\nabla^{2}}{2m}+e V_{eN}+A_0\right)\psi_{\alpha}+\cr &+&
\int\frac{{\rm d}{\bf x}}{8\pi e^2} \left({{\vec \nabla} A_0}\right)^2\ . 
\label{HnrQED_A0}
\eer
Here $\psi_{\alpha}({\bf x})$ is the electron field operator, $\alpha$ is the spin index, 
$e$ is the electron charge, $V_{eN}({\bf x},{\bf R}_I)$ is a pseudopotential, 
{\it i.e.}, an effective potential of ions at positions ${\bf R}_I$ felt by the valence electrons 
\cite{PhysRev.112.685,PhysRevB.25.7403}; $A_0({\bf x})$ 
is 
the scalar potential operator which mediates electron electrostatic interactions. 
Note that the $A_0$ terms can be integrated out leading to the standard two-body 
Coulomb interaction operator. 

The Kohn-Sham (KS) equation of the orbital-based DFT with a semi-local 
exchange-correlation functional, such as that by Perdew, Burke and Ernzerhof (PBE) \cite{PhysRevLett.77.3865}, is
\ber
&&\left(-\frac{\hbar^{2}}{2m}\nabla^{2}+e V_{eN} + e V_{KS}\right)\phi_{i}({\bf x})=\epsilon_{i}\phi_{i}({\bf x}),\cr
&&V_{KS}=V_H+V_{xc},~V_H({\bf x})=\int{\rm d}{\bf x}'\frac{e n({\bf x}')}{|{\bf x}-{\bf x}'|}, \cr
&& V_{xc}({\bf x})=\frac{\delta E_{xc}[n]}{e\delta n({\bf x})},
\vspace{-1.2cm}
\label{KSphi}
\eer 
where $\phi_{i}({\bf x}), \epsilon_{i}$ are the KS orbitals and eigenvalues, respectively,
and $E_{xc}[n]$ is the exchange-correlation functional, and $n({\bf x})$ is the 
ground state density of valence electrons \cite{PhysRev.140.A1133,RevModPhys.80.3,PhysRevLett.77.3865}. 
In general, the state label $i$ may include band number, lattice wave vector and spin label. 
But in this work we only consider spin-symmetric aperiodic systems so that 
$\phi_{i\uparrow}({\bf x})=\phi_{i\downarrow}({\bf x})\equiv \phi_{i}({\bf x})$. 
Extending this approach to the case of a periodic and/or spin-polarized system
is straightforward. 
 
In order to utilize electronic structure information from the DFT output 
we introduce ${\rm a}_{i\alpha},$ which is the annihilation operator of a Fermion in the 
KS state $\ket{i,\alpha},$ 
so that
$\psi_{\alpha}({\bf x})=\sum_i\phi_{i}({\bf x}){\rm a}_{i\alpha}$~(see, {\it e.g.}, \cite{AGD, FW}).
In terms of ${\rm a}_{i\alpha}$ the imaginary-time (Euclidean) action corresponding 
to the Hamiltonian (\ref{HnrQED_A0}) is
\begin{widetext}
\ber
{S}_{E}=\int_{0}^{\beta}{\rm d}\tau \left(
\sum_{ij\alpha}
a^{\dagger}_{i\alpha}(\tau)
\left[\left(\partial_{\tau}+\epsilon_i-\mu\right)\delta_{ij} - V^{xc}_{ij} + {\rm i} {A_{ij}}\right]a_{j\alpha}(\tau)-{\rm i}\int{\rm d}{\bf x} A_0 n({\bf x})
+\int\frac{{\rm d}{\bf x}}{8\pi e^2} \left({{\vec \nabla} A_0}\right)^2\right),
\label{SE}
\eer
\end{widetext}
where $~\beta={1}/{T_e},~T_e$ is the electronic temperature in energy units used in the simulation, and
\ber
A_{ij}(\tau)&=&\int{\rm d}{\bf x}\phi^*_{i}({\bf x})A_0(\tau,{\bf x})\phi_{j}({\bf x}), \\ V^{xc}_{ij}&=&e\int{\rm d}{\bf x}\phi^{*}_i(\xx)V^{xc}(\xx)\phi_j(\xx)\ .
\label{AijVxcij}
\eer
The spatial integration is over the simulation box volume; KS indices 
$i$ and $j$ vary over the range of KS orbitals included in the simulation - the ``active window". 
Fermion fields $a_{i\alpha}(\tau)$ are anti-periodic $a_i(\tau)=-a_i(\tau+\beta),$ 
while the $A_0$-fields are periodic in time. The grand canonical 
chemical potential $\mu$ is set at the mid-gap which ensures charge neutrality. 
In order to obtain (\ref{SE}) we added and subtracted 
$e V_{KS}({\bf x})$ from (\ref{KSphi}) to the electron lagrangian,
and then used ${\nabla}^2 V_{H}=-4 \pi e n({\bf x})$ 
combined with the shift-invariance of the functional integration over $A_0.$
The statistical sum is given by
\ber
Z(\mu,T_e) =  \dsp \int {\rm D} A{\rm D} a_i {\rm D} a^{\dagger}_i\exp \left(-{S}_{E}\right).
\label{Z}
\eer
The advantage of using KS orbital basis is that these states 
approximate binding of electrons to the ions and some of the electron 
interactions. 
Also, since KS states are labeled by their energy, it is straightforward 
to only include few KS states near the Fermi level that are relevant to the description 
of low-energy excitations, which is more efficient than using 
spatial grid covering the whole simulation cell. Thus, in this 
approach low-energy excitations of a nanoparticle are described by 
the KS quasiparticles subject to the static potential $V^{xc}_{ij}$ 
and interacting via $A_{ij}(\tau)$ exchanges. 
Strictly speaking, one should proceed using the basis 
of $\epsilon_i\delta_{ij} - V^{xc}_{ij}$ eigenstates.
However, setting $V^{xc}_{ij} = V^{xc}_{ii}\delta_{ij}$ 
is an approximation often made in the GW method \cite{RevModPhys.74.601, GovoniGW}. Also,  
for the systems we have simulated and for the range of KS states 
included, $V^{xc}_{ij}$ is strongly dominated by its diagonal entries.
So, 
here we also approximate $V^{xc}_{ij} = V^{xc}_{ii}\delta_{ij}.$

Next we perform Grassmann integration over the Fermion variables and expand the resulting 
action ${S}(A_0)$ in powers of $A_0(\tau,{\bf x}).$ The terms linear in $A_0$ 
cancel, which reflects the system's overall charge neutrality. Retaining 
the leading non-vanishing term in the expansion yields the following action
\ber
{S}_2=-\int \frac{{\rm d}\tau{\rm d}\xx}{8 \pi e^2}A_0 \nabla^2 A_0 - {\rm tr}{\cal M}{\cal A}{\cal M}{\cal A} + {\cal O}\left({A_0}^3\right),
\label{S2}
\eer
where 
\ber
{\cal M}^{-1}_{ij}(\tau)=\left(\partial_{\tau} + \tilde{\epsilon}_i\right)\delta_{ij},~\tilde{\epsilon}_i = \epsilon_i - \mu - V^{xc}_{ii},
\label{M0}
\eer 
is the inverse non-interacting propagator and ${\cal A}_{i \tau, j \tau^{'}}=A_{ij}(\tau)\delta_{\tau \tau^{'}}.$ 
The second term in~\eqref{S2} includes the random phase 
approximation (RPA) polarization insertion in the KS orbital basis. 
The approximate action~\eqref{S2} has been 
used in this work to simulate the nanoparticle electrons. But, in order to 
evaluate the statistical sum~\eqref{Z} numerically we define the action 
on a discretized space-time grid. 
The Lagrangian corresponding to the Hamiltonian in~\eqref{HnrQED_A0} is invariant 
under time-dependent, spatially uniform $U(1)$ gauge transformations
\ber
\psi^{'}(\tau)=e^{{\rm i} \Lambda(\tau)}\psi(\tau),~A_0^{'}=A_0 - \partial_{\tau} \Lambda(\tau),
\label{U1gauge}
\eer 
where $\Lambda(\tau)$ is a function of time \cite{PhysRevB.79.165425}. 
Note that the gauge field action - the last term in~\eqref{SE} - is invariant 
under (\ref{U1gauge}) \cite{PhysRevB.79.165425} \footnote{One recognizes that the approximate action \eqref{S2} is not gauge invariant. However, the Coulomb gauge is already assumed in
 the KS equation.}.

To generate 
$A_0$-field configurations we have used the 
frequency representation and a spatial grid, {where the 
polarization term was expressed as a function of $\xx,\yy$ using $\phi_{j}({\xx}).$ } 
In this case $A_0(\omega,\xx)$ can be used instead of 
the link variables required on a Euclidean time lattice \cite{PhysRevD.10.2445}.
However, a $\tau$-KS lattice, where ${\{}A_0(\omega,\xx){\}}$ have 
been converted 
to the link variables, has been used to compute the observables. 
While more computationally expensive than a $\tau$-KS 
lattice, the $\omega\xx$ basis is used
since 1)~numerical cancellation of the ``tadpole" terms in 
a simulation requires perfect representation of the time 
derivative operator; 2)~the Laplacian in the gauge 
field action cannot be represented accurately with the few KS orbitals included 
in a reasonably sized active window. Then, the action is
\ber
{S}_2(A_0)=-\sum_{k \geq 0}\sum_{\xx, \yy}A_0^{*}(\omega_k,\xx) {\cal S}_k(\xx, \yy) A_0(\omega_k,\yy),
\label{S2Sxy}
\eer
where $\omega_k=2 \pi k T_e$ is a bosonic Matsubara frequency, and
\begin{widetext}
\ber
{\cal S}_k(\xx, \yy)&=&\frac{a_x a_y a_z T_e}{8 \pi e^2}\nabla^2_{\xx, \yy}+2(a_x a_y a_z)^2\sum_{ij}\phi^*_{i}(\xx)\phi_{j}({\xx})\phi^*_{j}(\yy)\phi_{i}(\yy)\frac{n_j-n_i}
{2 \pi {\rm i} k -\beta{\tilde{\epsilon}}_{j}+\beta{\tilde{\epsilon}}_{i}},\cr
\nabla^2_{\xx, \yy}&=&\sum_{l=1}^3\left(\nabla_l^2\right)_{\xx, \yy},~\nabla_l^2f(\xx)=\frac{1}{a_l^2}\left(f\left(\xx+a_l\hat{l}\right)-2f(\xx)+f\left(\xx-a_l\hat{l}\right)\right)+{\cal O}(a_l^2),
\label{Mxy}
\eer
\end{widetext}
\vspace{-14mm}
where $a_l,~l={x,y,z},$ are the lattice spacings and $n_i=\left({\rm exp}\left[\beta{\tilde{\epsilon}}_{i}\right]+1\right)^{-1}.$
Gauge invariance requires that each included $A(\omega)$ 
is coupled to at least one pair of fermion modes $a^{\dagger}(\omega_1),~a(\omega_2)$ with 
$\omega=\omega_2-\omega_1.$ Therefore, in the simulations where a finite 
frequency cutoff was used the expression for ${\cal S}_k(\xx, \yy)$ 
was modified accordingly. 

Importantly, ${\cal S}_k( \xx, \yy)$ from (\ref{Mxy}) is 
real and symmetric under $\xx \leftrightarrow \yy,$ 
which is due to the basic properties of KS eigenfunctions 
(see, {\it e.g.}, \cite{landau1981quantum}).
Then the action ${S}_2(A_0)$ from (\ref{S2Sxy}) is 
{\it non-negative},
which suggests that 
in this approach 
meaningful simulations are possible without resorting to sign-suppression techniques 
(see, {\it e.g.}, \cite{Fodor:2001au,PhysRevLett.102.131601}). The size of neglected 
higher order terms in the action expansion will be discussed later.

The action (\ref{S2Sxy}) is quadratic in $A_0$. Therefore, 
importance sampling is done by diagonalizing ${\cal S}(\omega_k,\xx, \yy)$ 
for each $\omega_k$ which results in 
\ber
{S}_2(A_0)=\sum_{k \geq 0} \sum_{i=1}^{N_x-1} \frac{\lambda_i(\omega_k)}{2} \left(u(\omega_k)_i^2+{v}(\omega_k)_i^2\right), 
\label{Sdiag}
\eer
where $\lambda_i(\omega_k)>0, {v}(0)_i \equiv 0,~N_x$ is the number 
of spatial grid points. Then one generates random values for the $u(\omega_k)_i,~{v}(\omega_k)_i$ 
variables that are distributed normally according to the corresponding 
eigenvalue $\lambda_i(\omega_k),$ and changes the basis to get the 
$A_0(\tau,\xx)$ configurations. For all the nanocrystals simulated 
here ${\cal N} \simeq 10^3$ configurations have been found to be sufficient to 
obtain statistically significant results.

The observables considered here are the electron and hole quasi-particle 
energies, which have been extracted from two-point propagators,
which are {the averages of the matrix elements of the propagator matrix 
corresponding to the state $i$ and time slices $\tau_1$ and $\tau_2,$}
 in the long-time limit.
{For a particle state $i > HO,$ where $HO$ labels the 
highest occupied orbital, it is $E_g \beta \gg E_g\tau \gg 1,$
where $E_g$ is the quasiparticle gap.}
Then
\begin{align}
&{\rm M}_i(A_0)(\tau_2,\tau_1)=\left(\delta\left[{\rm D}+\tilde{\epsilon}_i\right]\right)^{-1}|_{i\tau_2,i\tau_1},\nonumber\\
&{\rm D}_{ij}f(\tau) = \frac{1}{\delta}\left(f(\tau+\delta)-e^{-{\rm i} \delta  A_{ij}(\tau)}f(\tau)\right),\nonumber\\
&{\rm M}_i(\tau,0)=\<{{\rm M}_i(A_0)}\> =   \sum_{k=1}^{{\cal N}}\frac{{\rm M}_i(A^k_0)}{\cal N}\rightarrow {}Ce^{-E_i\tau}\ ,
\label{MA}
\end{align}
where $\delta=\beta/N_t,~N_t$ is the number of time grid points, $A_{ij}(\tau)$ 
is defined in (\ref{AijVxcij}), $\tilde{\epsilon}_i$ is defined in 
(\ref{M0}). The last line in 
\eqref{MA} shows the behavior of the correlator at large times on the 
fully interacting excitation energy $E_i$ ($C$ is an irrelevant coefficient).  
We fit for this energy and estimate the statistical error of $E_i$ via the 
bootstrapping technique~\cite{kunsch1989}. A full error analysis, which would 
include an assessment of systematic errors due to, \emph{e.g.}, fit-window sizes 
and fitting functions, finite lattice spacings, {\it etc.}, is left 
to future work, though we do not expect the uncertainties quoted 
here to change significantly. We have used time grids with $\delta=0.025~eV^{-1}$ 
for all systems after checking that KS energies can be accurately extracted from 
the propagators in the non-interacting cases. {Frequency cutoffs were chosen so that $\omega_{max}\delta \simeq 1.$} The procedure to extract hole excitation
energies is analogous.

\begin{figure}[t]
\centerline{\includegraphics[width=.7\columnwidth]{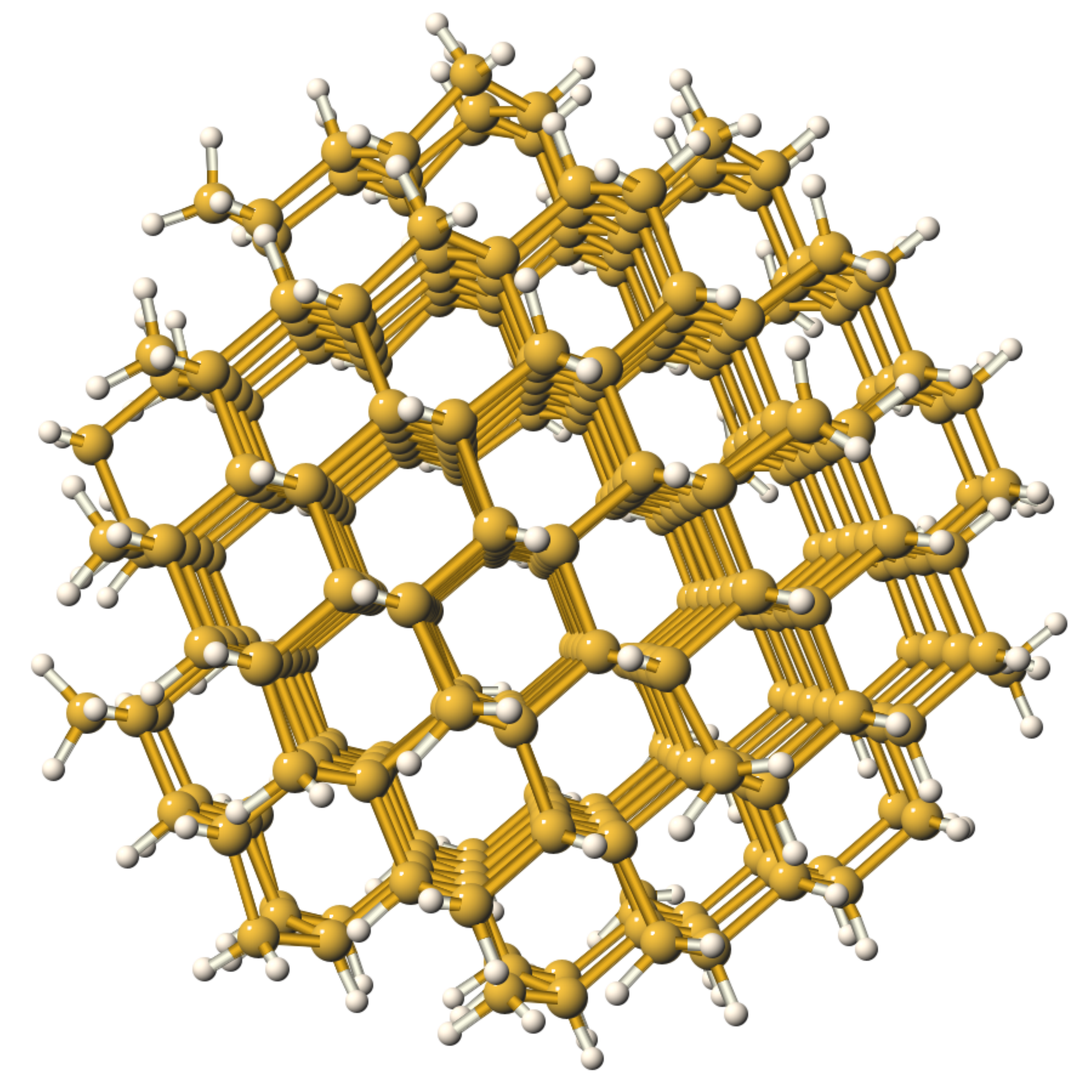}}
\hspace{-.1in}
\caption{Atomistic model of $Si_{293}H_{172}.$ The smaller (white) surface atoms are the hydrogens.}
\label{fig1}
\end{figure}

DFT simulations of the atomistic models of the nanocrystals 
(such as $Si_{293}H_{172}$ shown in Fig. \ref{fig1}), including geometry 
relaxation, have been done using Quantum Espresso DFT program  
with the PBE exchange-correlation functional \cite{0953-8984-21-39-395502}. 
Norm-conserving pseudopotentials \cite{PhysRevLett.43.1494} have been used 
ensuring that the KS orbitals 
are orthonormal.
Kinetic energy cutoff, which determines lattice spacings 
$a_i,~i=x,y,z,$ has been set to $340.1~eV,$ which is 
the same as in \cite{GovoniGW}, resulting in $a_i\simeq 0.05~nm.$ Models of the nanocrystals 
ranging in size from $1.0$ to $2.4~nm$ were placed in the periodic 
cubic simulation boxes with about $1~nm$ of vacuum between 
the surfaces in order to prevent spurious interactions between periodic 
images. The number of KS orbitals included in the simulations
have been chosen so that 
$\epsilon_{i_{max}}-\epsilon_{HO}\simeq\epsilon_{HO+1}-\epsilon_{i_{min}}\geq 1.5 E^{PBE}_g,$ 
where $i_{max},~i_{min}$ are the highest and the lowest included KS orbital labels and $E^{PBE}_g=\epsilon_{HO+1}-\epsilon_{HO}$ is
the non-interacting gap. The corresponding number of states above/below Fermi level included in the simulation varied from 36 to 96 as the system's size increased.

For the nanoparticles considered in this work $V^{xc}_{ii}$ shifts are 
sizable and tend to 
shrink the non-interacting gap 
$E^{PBE}_g$. 
This would require lowering $T_e$ in order to maintain $E^{PBE}_g \gg T_e$ 
required for the tadpole term cancellation, which would 
significantly increase the computational expense. 
So, here we have treated $V^{xc}_{ii}$ as 
self-energy corrections, {\it i.e.}, we have simulated with 
$\tilde{\epsilon}_i = \epsilon_i - \mu$ and then subtracted 
$V^{xc}_{ii}$ from the resulting single-particle energies. 

{\begin{table}
\raisebox{0.00001\totalheight}{\begin{tabular}{|c|c|c|c|c|}\hline
\textbf{} & $Si_{35}H_{36}$ & $Si_{87}H_{76}$ & $Si_{147}H_{100}$& $Si_{293}H_{172}$ \\ \hline
\textbf{$T_e, ~eV$} & $0.5$ &  $0.5$ & $0.4$ & $0.3$ \\ \hline
\textbf{$N_{e}$} & $176$ &  $424$ & $688$ & $1344$ \\ \hline
\textbf{$\<{\tilde s}_2\>$} & $74.1(5)$ &  $76.2(3)$ & $315.4(3.4)$ & $167.2(3)$ \\ \hline
\textbf{$Re\<{s}\>$} & $64.0(3)$ &  $70.0(2)$ & $287.9(2.1)$ & $162.2(2)$\\ \hline
\textbf{$Im\<{s}\>$} & $0.1(2)$ &  $-0.1(1)$ & $0.2(2)$ & $-0.1(2)$ \\ \hline
\textbf{{$\<{\tilde s}_2/|{s}|\>$}} & { $1.155(1)$} &  { $1.0865(7)$} & ${ 1.073(3)}$ & ${ 1.0294(3)}$\\ \hline
\textbf{{$\<{\rm tan} (\varphi)\>$}} & {$0.002(3)$} & {$(-0.6 \pm 2.0)\cdot 10^{-3}$} & {$0.0006(8)$} & {$-0.0005(12)$}\\ \hline
\end{tabular}}
\label{S}
\caption{{Both $s$ and ${\tilde s}_2$ dimensionless; ${s}=|s|e^{i \varphi}$. Numbers in parentheses represent statistical errors. $N_e$ is the valence electron number.}} 
\end{table}}

\begin{figure}[t]
\hspace{-.5in}
\centerline{\includegraphics[width=1.1\columnwidth]{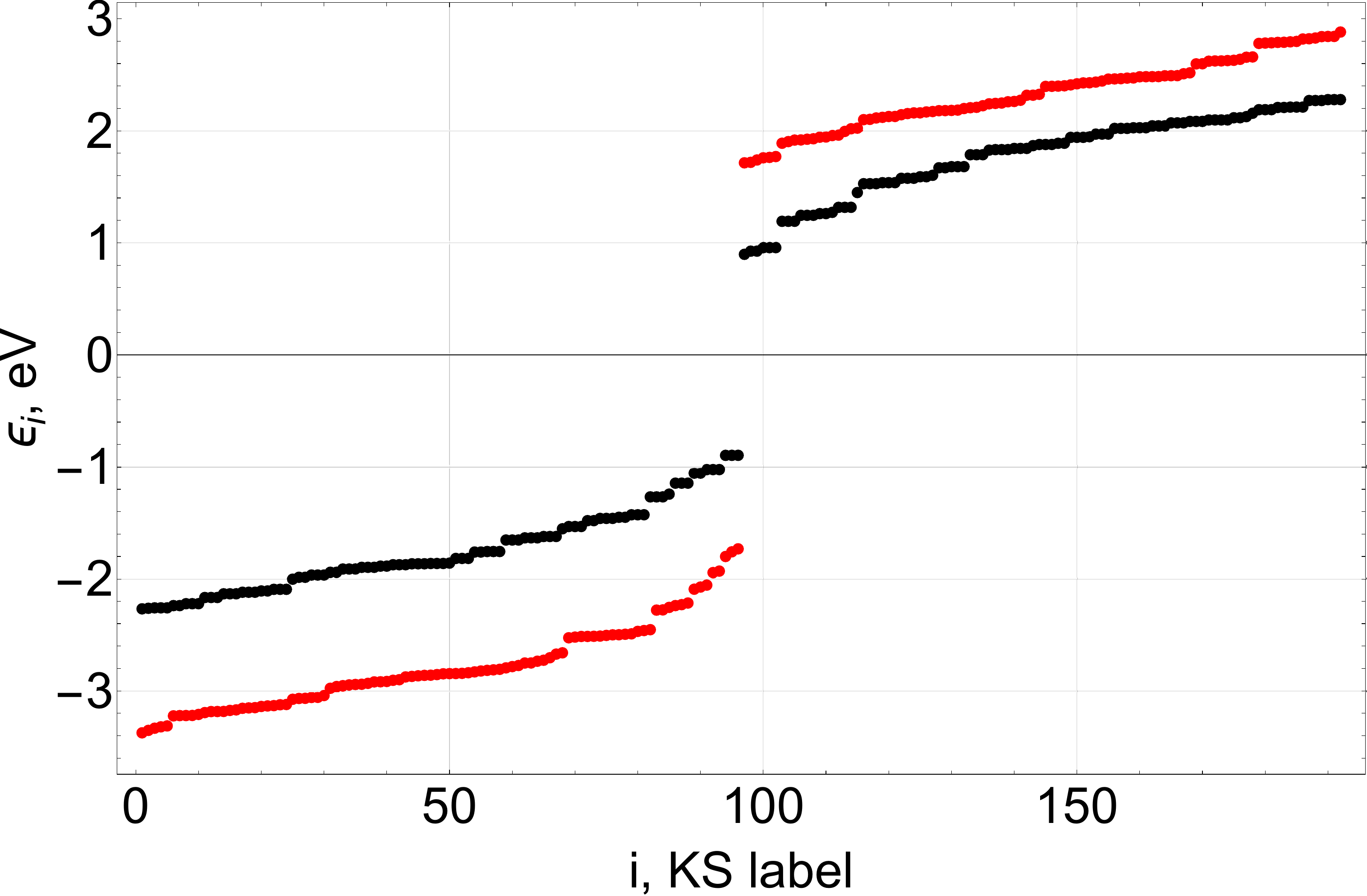} }
\hspace{-.5in}
\caption{Interacting quasi-particle energies (red) for $Si_{293}H_{172}$. Black points are the KS eigenvalues. Error bars are too small to be seen at this scale.}
\label{fig2}
\end{figure}

The calculation results are shown in Tables I and II. 
In order to check the size of the terms neglected in the approximate action (\ref{S2Sxy}) we used the 
A-configurations generated with (\ref{S2Sxy}) to compute
\ber
{s}&=&{\rm tr}~{\rm log} \left(\frac{{\cal M}^{-1} 
+ {\rm i} {\cal A}}{{\cal M}^{-1}}\right)- {\rm i}~{\rm tr}\left({\cal M}{\cal A}\right),\cr
{\tilde s}_2&=&\frac{1}{2}{\rm tr}\left({\cal M}{\cal A}{\cal M}{\cal A}\right).
\label{S2S}
\eer
While ${\tilde s}_2$ is only the leading non-vanishing term in 
the expansion of $s,$ we have found that in all cases $\<{\tilde s}_2/|{s}|\>$ is close 
to one. (See Table I.) This suggests that the full Fermion action can be reasonably approximated by just 
the leading term. 
The average of the action's phase $\varphi,$ where $s=|s|e^{i \varphi},$ is close to zero, 
which suggests that the sign problem in these systems is mild. Shown in Table II are the quasiparticle gap predictions in 
these nanocrystals. Our results agree with the results of high-precision 
calculations using $G_0W_0$ for the same nanocrystals \cite{GovoniGW}. Low-energy 
particle and hole levels in $Si_{293}H_{172}$ are shown in Fig. \ref{fig2}. 
\hspace{-.0025in}
{\begin{table}
\raisebox{0.00001\totalheight}{\begin{tabular}{|c|c|c|c|c|}\hline
\textbf{} & $Si_{35}H_{36}$ & $Si_{87}H_{76}$ & $Si_{147}H_{100}$ & $Si_{293}H_{172}$\\ \hline
\textbf{$E_g^{PBE}$} & 3.51 &  2.59 & 2.29 & 1.79 \\ \hline
\textbf{$E_g^{QP}$} & { $6.29(9)$} & {$4.76(8)$} & { $4.22(6)$} & {$3.45(3)$}\\ \hline
\textbf{$E_g^{G_0W_0}$} & { 6.29 } & { 4.77 } & { 4.21 } & { 3.46}\\ \hline
\end{tabular}}
\label{Egresults}
\caption{All entries are in eV. $E_g\equiv\epsilon_{HO+1}-\epsilon_{HO}$, 
is the PBE gap; the interacting gap is $E^{QP}_g=E_{p}^{min}-E_{h}^{max}.$ ${G_0W_0}$ results are from \cite{GovoniGW}.} 
\end{table}}

In conclusion, we have performed initial steps toward development 
of a first-principles high-precision Monte Carlo technique for the 
excited states of a semiconductor nanoparticle, which utilizes 
KS orbital basis in the imaginary-time functional integral for the 
electrons.  
We find that approximating Fermion action with the leading order 
RPA polarization term in the expansion in powers of $A_0$ leads to 
a positive definite integrand in the statistical sum and that it is 
a reasonable approximation to the full action; $\<{\tilde s}_2/|{s}|\>-1$ can be viewed
as a source of systematic error. So, our 
results suggest that in this approach 
these systems have only a mild Fermion sign problem. 
Obvious improvements to our approximate calculations would be 
to use the full action 
instead of our quadratic approximation in~\eqref{S2}, which 
could be done via re-weighting \cite{doi:10.1063/1.1729945,PhysRevLett.61.2635} 
or more advanced sampling techniques ({\it e.g.},
hybrid Monte Carlo \cite{Duane:1987de}).  
Work to develop technique for other excited states, such as 
excitons and trions, is in progress.

The authors acknowledge use of computational resources at
the Center for Computationally Assisted Science and
Technology (CCAST) at North Dakota State University. 
 TL acknowledges financial support from the Deutsche Forschungsgemeinschaft (Sino-German CRC 110). VVK acknowledges support by the Department of Energy National Nuclear Security Administration under Award No. DE-NA0003856.
\bibliography{dftNSF2013}
%

\end{document}